\documentclass[twocolumn, balance, trackchanges]{aastex631}

\usepackage{flushend}
\submitjournal{Astrophysical Journal}
\received{March 5th, 2023}
\revised{April 3rd, 2023}
\accepted{April 6th, 2023}
\shorttitle{Prompt Detection of Fast Optical Bursts}
\shortauthors{Megias Homar, Meyers and Kahn}

\graphicspath{{./}{figures/}}

\begin{document}

\title{Prompt Detection of Fast Optical Bursts with the Vera C. Rubin Observatory}

\author[0000-0001-6013-1131]{Guillem Megias Homar}
\affiliation{Department of Aeronautics and Astronautics, Stanford University, Stanford, CA 94305, USA}
\affiliation{Kavli Institute for Particle Astrophysics and Cosmology, Stanford University, Stanford, CA 94309, USA}
\affiliation{SLAC National Accelerator Laboratory, Menlo Park, CA 94025, USA}
\affiliation{Vera C. Rubin Observatory, Colina El Pino Casilla 603, La Serena, Chile}

\author[0000-0002-2308-4230]{Joshua M. Meyers}
\affiliation{Vera C. Rubin Observatory, Colina El Pino Casilla 603, La Serena, Chile}
\affiliation{Lawrence Livermore National Laboratory, Livermore, CA 94550, USA}

\author[0000-0003-4833-9137]{Steven M. Kahn}
\affiliation{Kavli Institute for Particle Astrophysics and Cosmology, Stanford University, Stanford, CA 94309, USA}
\affiliation{SLAC National Accelerator Laboratory, Menlo Park, CA 94025, USA}
\affiliation{Vera C. Rubin Observatory, Colina El Pino Casilla 603, La Serena, Chile}
\affiliation{Departments of Physics and Astronomy, University of California, Berkeley, Berkeley, CA 94720, USA}

\correspondingauthor{Guillem Megias Homar}
\email{gmegias@stanford.edu}

\begin{abstract}
 The transient optical sky has remained largely unexplored on very short timescales. While there have been some experiments searching for optical transients from minutes to years, none have had the capability to distinguish millisecond Fast Optical Bursts (FOB). Such very fast transients could be the optical counterparts of Fast Radio Bursts (FRB), the prompt emission from $\gamma$‐Ray Bursts (GRB), or other previously unknown phenomena. Here, we investigate a novel approach to the serendipitous detection of FOBs, which relies on searching for anomalous spatial images.  In particular, due to their short duration, the seeing distorted images of FOBs should look characteristically different than those of steady sources in a standard optical exposure of finite duration. We apply this idea to simulated observations with the Vera C. Rubin Observatory, produced by tracing individual photons through a turbulent atmosphere, and down through the optics and camera of the Rubin telescope. We compare these simulated images to steady-source star simulations in 15 s integrations, the nominal Rubin exposure time. We report the classification accuracy results of a Neural Network classifier for distinguishing FOBs from steady sources. From this classifier, we derive constraints in duration-intensity parameter space for unambiguously identifying FOBs in Rubin observations. We conclude with estimates of the total number of detections of FOB counterparts to FRBs expected during the 10-year Rubin Legacy Survey of Space and Time (LSST).
\end{abstract}

\keywords{Optical bursts (1164) --- Burst astrophysics (187) --- Radio transient sources (2008) --- High energy astrophysics (739)}

\section{Introduction}

Time-domain astronomy has attracted enormous interest in the community in recent years, as new facilities come on line that dramatically improve our capability for transient detection and source characterization. While much of the attention has been focused on traditional classes of variable stars and supernova explosions, the study of very fast transients has also gained substantial momentum. That field began with the serendipitous detection of the first $\gamma$-Ray Bursts (GRBs) in 1967 \citep{Klebesadel1973} and the subsequent searches for counterparts at optical, X-ray, infrared and radio wavelengths \citep{Paczynski1993, Panaitescu1998, Gehrels2004}. More recently, the discovery of the first Fast Radio Burst (FRB) in 2007 \citep{Lorimer2007}, with a duration less than 5 milliseconds, has led to a new era in fast transient astronomy. 

Despite the renewed interest, however instrumental limitations make the search for Fast Optical Bursts (FOBs) very challenging, and, as a consequence, the transient optical sky on very short timescales has remained largely unexplored. There have been many searches for optical transients with durations down to minute timescales: the Palomar Transient Factory on the 48 inch Samuel Oschin telescope at Palomar Observatory \citep{Law2009}, the ATLAS optical domain survey that reported the first discovery of a Fast Blue Optical Transient (FBOT) \citep{Prentice2018}, the Zwicky Transient Facility that reported a larger sample of FBOTs \citep{ho2021photometric} and has been used for the search of dirty fireballs \citep{Ho_2022}, the Dark Energy Survey \citep{Andreoni2019}, but none of these had the capability to recognize millisecond optical bursts. The very short timescales involved, and the presumed rarity of such events preclude most conventional approaches to such studies. Currently, only the Tomo-e Gozen wide-field high-speed camera on the Kiso 1.0-m Schmidt telescope capable of 0.2 ms accuracy \citep{sako2018} would be able to directly detect such events using millisecond-long exposures.

FOBs could have multiple origins: optical counterparts of FRBs, prompt emission from fast GRBs or other previously unknown phenomena. The highly energetic emission mechanisms that give rise to FRBs  \citep{Lyutikov2017, Petroff2019, Zhang2020} should also produce both prompt and afterglow emission in the optical band \citep{Yang2019}. For GRBs, afterglows have been widely observed and studied for many years \citep{Reem1999, Akerlof2000}, but the detection of prompt emission is far less common. In both cases, the discovery of FOB counterparts would further elucidate our understanding of the progenitor and environment of such exotic events. 

The conjunction of the CHIME/FRB Collaboration with other experiments has now yielded a substantial increase in FRB observational data \citep{Amiri2021, Caleb2017, Bhandari2018}. However, the extremely short duration precludes multi-wavelength followup via traditional target-of-opportunity campaigns. The exception involves the relatively small fraction of FRBs that are known to be repeating \citep{Caleb2018}, for which simultaneous observing campaigns in multiple bands can be coordinated \citep{Hardy2017, Eftekhari2018}. To date these have resulted in only a single detection of an optical transient spatially coincident with the repeating FRB180916B, which had an association probability of 0.04\% \citep{Li2022}. As for GRBs, the majority of searches for FRB counterparts have focused on afterglow multi-wavelength counterparts \citep{Petroff2014, Nunez2021, Marnoch2020} as opposed to prompt emission.

The principal obstacle to the search for FOBs lies in the scarcity of such events.  For uniformly distributed sources, the detection rate expected for bursts of a given fluence is proportional to the volume of space that can be surveyed to that fluence level at any given time.  That volume is in turn proportional to the product of the cube of the diameter of the primary mirror and the field of view of the camera. The upcoming Vera C. Rubin Observatory, with its 8.4 m Simonyi Survey Telescope and $9.6 \ \mathrm{deg^2}$ LSST Camera will provide a factor of nearly 20 improvement in this parameter over all existing and planned facilities operating in the optical band. As such, the Rubin Observatory offers the possibility of a dramatic advance in our ability to search for FOBs, if very short duration events can be uniquely identified in the images.

Here, we investigate a novel approach to the serendipitous detection of FOBs through the fact that their seeing distorted images should look characteristically different than those of steady sources in a standard optical exposure of finite duration.  In particular, for a steady source, the Point Spread Function (PSF) due to seeing involves an average over the distortions due to atmospheric turbulent structures that transit over the aperture while the image is being collected. In contrast, a fast optical flash of very short duration will exhibit distortions due to a much more limited patch of sky given by the projection of the primary aperture on the turbulent layers. As such, the PSF will exhibit structure at higher spatial frequencies. 
\begin{figure*}
    \centering
    \plotone{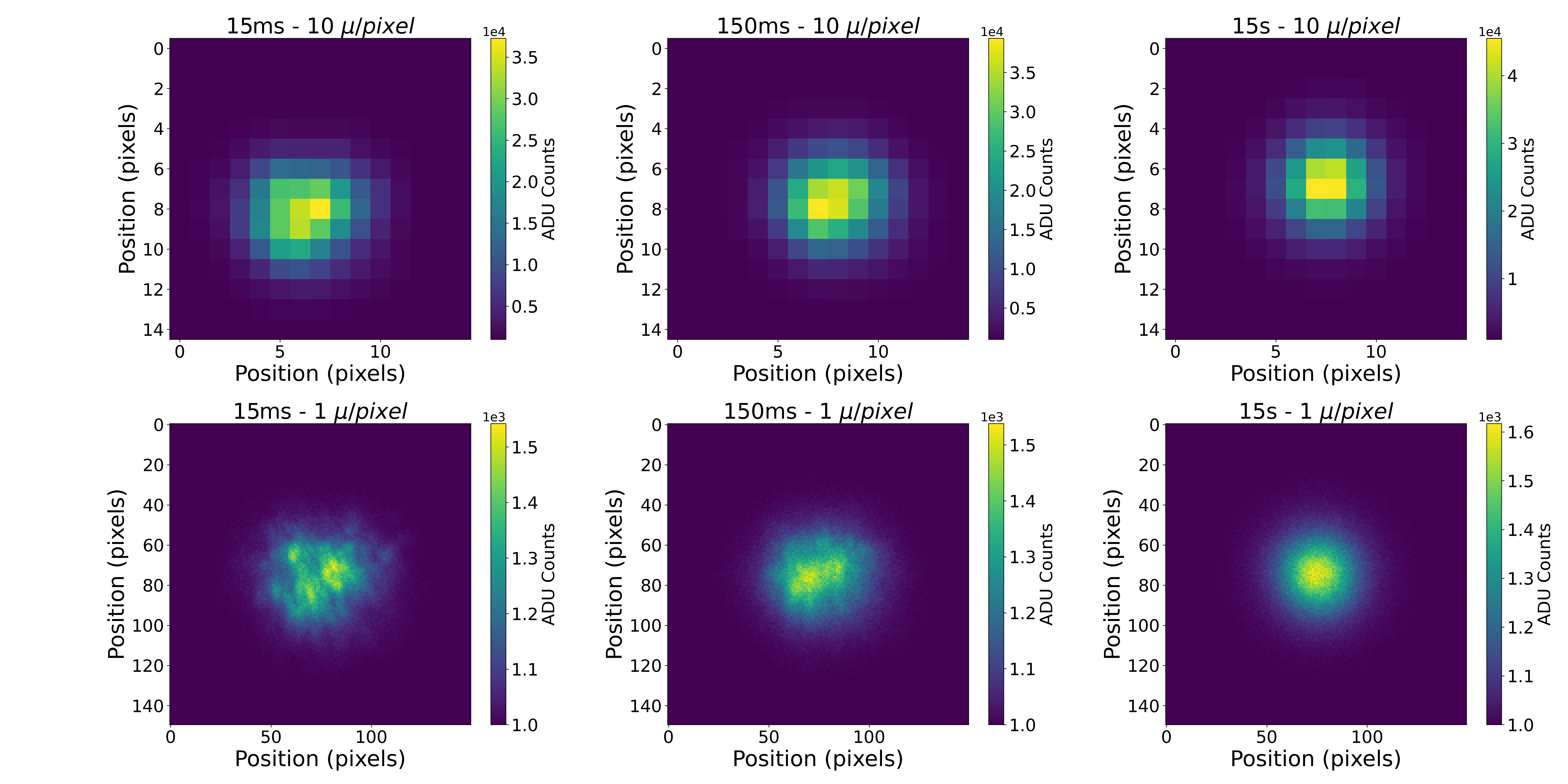}
    \caption{Set of simulated images. On the left, a 15 ms simulated FOB, on the middle a 150 ms FOB, and on the right a 15 s steady source. Top row shows the actual LSST Camera pixelization. Bottom row shows images with higher granularity.}
    \label{fig:simulation_sample}
\end{figure*}
In this work, we apply this idea to simulated observations from the Rubin Observatory. We simulate FOB observations by tracing photons both through an atmospheric model and through an optical model of the Rubin Observatory telescope. We compare these simulations to point-source star simulations of 15 s duration, the nominal Rubin exposure time. We then introduce a Neural Network classifier that discriminates between PSFs of optical flashes and steady sources. We report its performance and derive constraints in the duration-intensity parameter space for identifying FOBs in Rubin observations. Finally, we present an estimate of the total number of Rubin Observatory detections of FOB counterparts to FRBs expected during the 10-year Large Survey of Space and Time (LSST). For that estimate, we assume two different optical-to-radio fluence ratios. Our results reinforce the importance of building the scientific capability to identify FOBs into Rubin Observatory pipelines.

\newpage
\section{Methods}

\subsection{Simulation Model}
To generate images of fast optical flashes of different durations and intensities, we performed sequential photon by photon tracing that interfaced several different models. In particular, we used the Ellerbroek model \citep{Ellerbroek2002} to simulate the atmospheric structure by a series of frozen phase screens drifted by wind velocity vectors, with each screen covering the turbulence between two altitudes. We then generated the atmospheric PSF by convolving the PSF of the different screens averaged during the FOB duration. Because of the short burst duration, we cannot use the two-bounce raytrace approximation introduced by  \citep{Peterson2015}, since that relies on temporal averaging. Instead, we compute the full Fourier transform diffraction integral over the phase screen. For each burst we then convolve a square-shaped burst profile with the time-dependent atmospheric PSF, thus obtaining the FOB effective PSF. We sample this PSF with a discrete number of individual photons consistent with the burst fluence that we are modeling. The atmospheric seeing used in these simulations was of 0.67 arcsec of Full-Width Half Maximum (FWHM) and Fried parameter $r_0 = 0.11$ m. In general, larger values for seeing would reinforce the differences between FOBs and steady sources. For convenience, the wavelength of the photons is set to 754.06 nm, which corresponds to the central wavelength of the Rubin Observatory i-band, but the wavelength dependence is relatively weak for all of the six bands. We used the \texttt{galsim} python package \citep{galsim2016} for this PSF generation part of the simulation.

The photons are then traced through the optical model of the Rubin Observatory available in the \texttt{batoid} python package \citep{batoid2019}. The photons are finally traced into the LSST Camera sensor accumulating them into the CCD pixels using the \texttt{galsim} python package. This part took into account different camera physical effects and systematic errors. In particular, we included the Tree-Ring effect \citep{Park2020} and Brighter-Fatter effect \citep{Lage2017}. Once the images were obtained, the background level noise was added based on anticipated sky levels for the i-band of the Rubin Observatory \citep{ivezic2010}. All of the simulated images are 20 px x 20 px in size.

\subsection{Datasets}

Different datasets of FOB images were generated along with simulated 15 s steady-source observations. More precisely, we generated 2000 observation images for FOBs with durations from 5 ms to 1000 ms and fluences from $8.80 \ \mathrm{Jy \ ms}$ down to $0.23 \ \mathrm{Jy \ ms}$. Additionally, we introduced a dataset with 3000 cosmic ray images taken from real Rubin Observatory Commissioning Camera dark flats that were generated explicitly for this study.

\subsection{Neural Network classifier}

To discriminate between FOBs and steady sources, we trained a deep convolutional Neural Network (NN) on a training dataset of 4000 images, for ten different reference fluence values. We used labeled subsets of 2000 images of 15 s steady sources and 2000 images of 15 ms FOBs. All the images were normalized prior to their use, to base our classification exclusively on the PSF shape, regardless of its intensity. We used transfer learning and based our NN structure on the VGG-16 deep neural network \citep{Simonyan2014}. We trained the classification network for 20 epochs over 10,000 training samples in batches of 256 for each of the fluence reference levels. Throughout the training, we evaluated the network on the training set to track its performance. After training was complete, we confirmed that the network's score on the testing set was close to that obtained on the training set. We used the Pytorch Python package to implement and train the networks \citep{Pytorch2019}. 

The performance of our NN classifier was then measured by evaluating our classifier on simulated FOB images of different durations and steady-sources images at the same 15 s fluence. Since fluence is directly measured in the images, we apply the trained classifier for the specific measured fluence level. We define two different metrics to characterize the classifier performance: completeness and purity. The completeness is the percentage of input FOBs that are correctly classified by the NN. The purity is the percentage of events classified as FOBs that actually are FOBs.

\section{Results}
\subsection{Simulation Model}
As expected, our simulated images show a distinct difference in the PSF structure between short duration FOBs and 15 s steady sources, where 15 s is the Rubin Observatory nominal exposure. In Figure \ref{fig:simulation_sample}, we present a sample of these images, both at the actual LSST Camera pixelization, and at higher granularity, in order to better illustrate the PSF shape differences. It is apparent from the images that the short duration events exhibit a higher degree of speckling. Even with the loss of information that occurs when the photons are accumulated into pixels, the characteristic differences in PSF structure for the short bursts are preserved. This establishes the feasibility of building a PSF classifier that has the potential to distinguish FOBs. The simulated images in Figure \ref{fig:simulation_sample} correspond to $8.80 \ \mathrm{Jy \  ms}$ events.

\begin{figure}[ht!]

    \centering
    \epsscale{1.25}
    \plotone{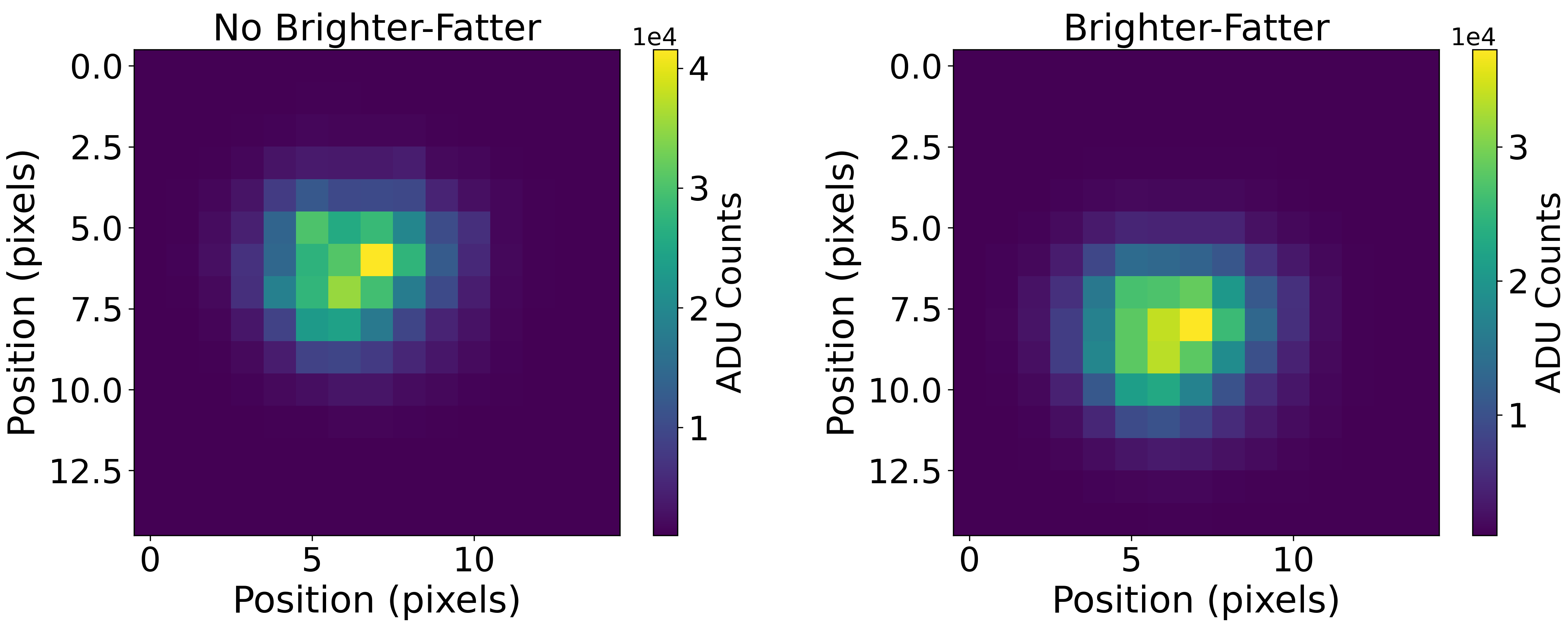}
    \caption{Comparison between a simulated image from a $15 \ \mathrm{ms}$ FOB in the LSST sensor including and excluding the Brighter-Fatter and tree rings effects.}
    \label{fig:brighterfatter}
\end{figure}

The Brighter-Fatter effect is a non-linear effect in CCDs in which a brighter point source produces a wider image, due to the mutual repulsion of photoelectrons as they begin to fill the pixel potential wells. It can play an important role for higher fluence events. Figure \ref{fig:brighterfatter} presents a comparison between a simulated image from a $15 \ \mathrm{ms}$ FOB in the LSST sensor including and excluding the Brighter-Fatter effect and related tree ring effects. The comparison is made for sources of $8.80 \ \mathrm{Jy \  ms}$, the highest fluence that we have considered. As can be seen, some of the high-frequency structure in the PSF is made more diffuse by the Brighter-Fatter effect, but not enough to prevent the discrimination between FOBs and steady sources.

\subsection{Neural Network classifier}
\begin{figure}[h!]
    \centering
    \epsscale{1.2}
    \plotone{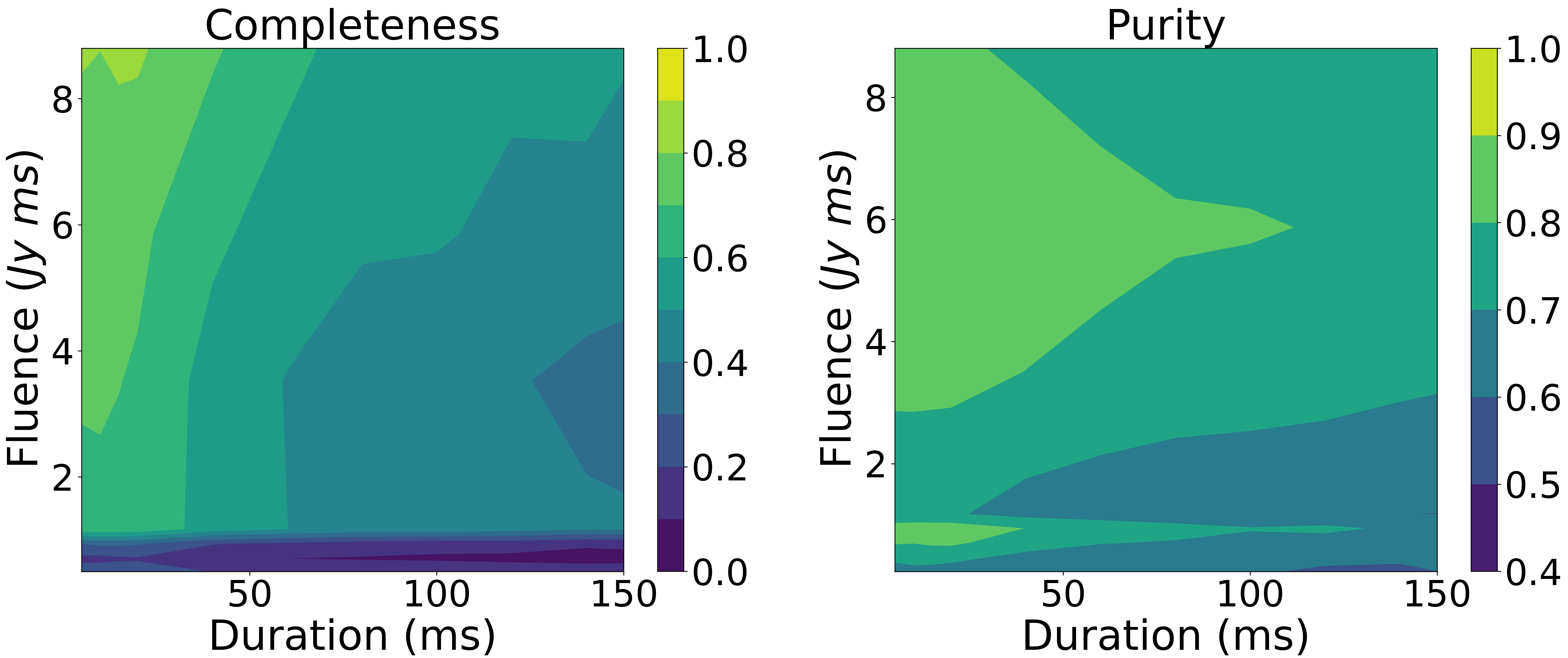}
    \caption{Completeness and purity of the NN classifier for datasets of FOBs with different durations and different fluences.}
    \label{fig:purity}
\end{figure}

In Figure \ref{fig:purity} we plot the completeness and purity of our NN classifier as a function of the duration and intensity of the burst. Very high completeness is obtained for short durations (from 5 ms to 30 ms) and a broad range of fluence levels (from $8.80 \ \mathrm{Jy \ ms}$ to  $3 \ \mathrm{Jy \ ms}$). We also obtain relatively high purity $>80\%$ in this range. That provides confidence that the approach we have described is capable of detecting and discriminating FOBs in Rubin images with reasonable reliability.  

For longer burst durations, the completeness decreases as the PSF starts to more closely resemble that of a steady source. We also find low completeness at low fluence, $< 1.17 \ \mathrm{Jy \ ms}$. In this range, photon counting statistic fluctuations mask the high frequency structure in the PSF for both FOBs and steady sources, so that it becomes impossible to tell them apart.  However the purity remains high, given the small fraction of images which are still classified as FOBs. In fact, slight tuning of the NN at low fluence leads the classifier to exhibit either high completeness and low purity, or, conversely, low completeness and high purity. The percentage of false positives remains constant around 10\% until it climbs for lower fluences, at which our discriminating power decreases.

We also trained a classifier to recognize cosmic rays and to discriminate them from both FOBs and steady sources. The purity of this classifier in discriminating cosmic rays is 95\%. Therefore, we do not expect the presence of cosmic rays in real Rubin images to preclude FOB detection.

\section{Discussion}
\begin{figure}[h!]
    \centering
    \epsscale{0.9}
    \plotone{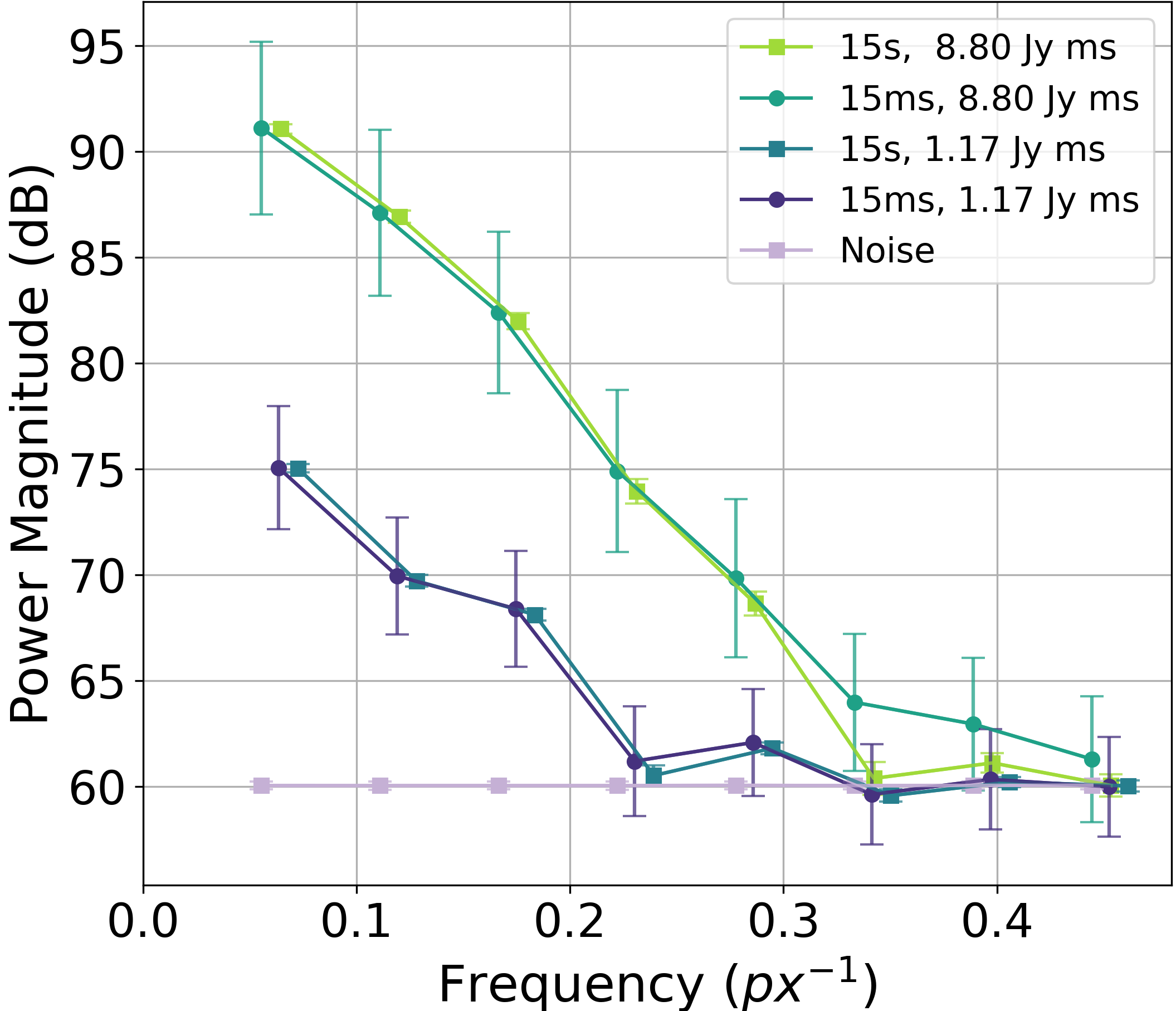}
    \caption{Comparison of the mean and variance of the power spectral density in the $x$-axis for 15 s and 15 ms FOB simulated images for the $8.8 \ \mathrm{Jy \ ms}$ and $1.1 \ \mathrm{Jy \ ms}$ fluence level. The power spectral density of the noise is shown in pale purple.}
    \label{fig:psd}
\end{figure}

If confirmed, the existence of Fast Optical Bursts would open a new channel of investigation in transient astronomy, and would help to elucidate the physical mechanisms that underlie such exotic phenomena. Our results show that if such bursts exist, we should be capable of detecting them. Indeed, the simulation results shown in Figure \ref{fig:simulation_sample} highlight the structural differences in the PSFs between FOBs and steady sources. We have devised a simple NN-based classifier that can detect and discriminate FOBs of very short duration. The strength of the deep convolutional NN classifier resides in the fact that it learns to differentiate between subtle differences in PSFs that cannot easily be characterized through simple statistical metrics. Nonetheless, a look into the power spectral density in the two directions of the image sheds light on these differences. In Figure \ref{fig:psd} we present the power spectral density in the $x$-direction for dataset FOB images of 15 ms and 15 s, we include both the highest considered fluence ($8.8 \ \mathrm{Jy \ ms}$) and the lowest ($1.1 \ \mathrm{Jy \ ms}$). While the higher fluence images show a characteristic difference at higher frequencies, the lower fluence power spectra are noise-dominated at those frequencies, making the difference indiscernible.

Despite the $> 80\%$ completeness for short burst durations, there is a noticeable decrease in completeness for durations of 30 ms and longer. That is a natural consequence of the averaging over turbulent structures in the atmosphere with time.  By 40 ms, the FOB images begin to closely resemble those of steady sources.  Further improvements in completeness might be achievable by explicitly training the NN on samples of longer duration FOBs.  In the current study, we trained the NN only with 15 ms FOBs.

The nominal Rubin operations plan involves two sequential exposures, each of 15 s duration, for each "visit" to a given field in the sky.  Alternatively, a single 30 s exposure could be performed, which would slightly improve the observing efficiency of the survey, since it would avoid the second 4 s readout of the CCDs.  While the longer single exposure would marginally increase the distinction between FOB and steady-source PSFs, the two-exposure mode is preferred for FOB detection, since it provides an easy way to develop a candidate set of short duration events through comparison of the two images.  Given such a candidate set for each visit, it would be straightforward to analyze those particular images for evidence of very short-duration burst structure.  Image subtraction of the two images would also provide a means of eliminating source crowding effects from galaxies and other steady sources in the field.  We strongly advocate that Rubin stay with its nominal operations plan to retain sensitivity for fast transient astronomy.

While the possibility of entirely new phenomena that give rise to FOBs is especially exciting, it is of interest to estimate how many such events might be detected with Rubin as prompt counterparts to FRBs.  The CHIME experiment measures an FRB all-sky rate at $600 \ \mathrm{MHz}$ of $525 \pm 30 \ \mathrm{sky^{-1} \ day^{-1}}$, to a fluence limit of $5 \ \mathrm{Jy \ ms}$.  We can extend this to calculate the FRB sky rates at different fluence limits using the fluence index $\alpha = -1.40 \pm 0.11$ \citep{Amiri2023, Amiri2021}. We then assume two possible representative optical-to-radio fluence ratios not motivated by any particular theoretical model ($\eta = 0.1$ and $\eta = 0.5$) to compute the estimated total number of FOBs that could be detected over the 10-year LSST Survey. From our simulations, we adopt the completeness factor associated with each fluence threshold considered, for the discrimination of FOBs in the Rubin images. The results of these calculations are shown in Table \ref{table}.

Note that the number of expected events is not huge, but is still large enough to yield a statistically meaningful sample.  The results do indicate, however, that a study like this is only feasible with Rubin given its factor 20 advantage over all other facilities.  It is therefore crucial for a classifier like the one we have developed to be incorporated into the Rubin science pipeline in order to enable a meaningful search for millisecond FOBs.

\begin{deluxetable}{l r r r r}
\tabletypesize{\small}
\tablecaption{\label{table} Estimated total FOB detections during the Large Survey of Space and Time (LSST) extrapolated from the FRB rates at 600 MHz obtained by CHIME.}
\tablecolumns{5}
\tablehead{
\colhead{Fluence} & \colhead{Completeness} & \colhead{FRB rate}  & \colhead{$\eta$} & \colhead{Survey FOBs} \\ \colhead{\footnotesize{(Jy ms)}} &  & \colhead{\footnotesize{(events/sky/day)}} &  & \colhead{\footnotesize{(events)}}}
\startdata
    $\geq$ 2 &  68.05 \% & 1893 $\pm$ 108 & 0.1 & 26.03 \\
      &  &   & 0.5 & 130.15 \\
    $\geq$ 4 &  70.8 \% & 717 $\pm$ 41 & 0.1 & 10.26 \\
      &  &  & 0.5 & 51.31 \\
    $\geq$ 6 &  74.4 \% & 406 $\pm$ 23 & 0.1 & 6.11 \\
      &  &   & 0.5 & 30.56 \\
    $\geq$ 8 &  80.9 \% & 271 $\pm$ 15 & 0.1 & 4.44 \\
      & & & 0.5 & 22.2
\enddata
\end{deluxetable}

\section{Conclusions and Future Work}
We have introduced a novel approach to detecting Fast Optical Bursts. The existence of FOBs would bring invaluable new information to the study of fast transient events, either FRBs, GRBs or other unknown phenomena. We simulated the PSFs of different FOBs using a Rubin Observatory simulation model and reported a successful approach for discriminating them from steady sources. 

In summary:  (1) Our simulation model showed that PSF structures for short‐duration FOBs are distinctively different than those of steady sources. (2) The combination of high completeness ($\mathbf{> 70\%}$) and high purity for FOBs above 3 Jy ms and shorter than 30 ms, make the certainty of our classification algorithm reliable. (3) The estimated FOB rates for FRB counterparts indicate that a study like this is only possible with Rubin, not with any of the smaller precursor surveys. 

There are a variety of ways in which our work could be extended. First and foremost, our results make a case for building the scientific capability of detecting these bursts into the Rubin Observatory scientific pipelines. The integration and refinement of our classifier into the Rubin pipelines, and the introduction of FOB source injection prior to first light are the main activities to be pursued.

\begin{acknowledgements}
    This material is based upon work supported in part by the National Science Foundation through Cooperative Agreement AST-1258333 and Cooperative Support Agreement AST1836783 managed by the Association of Universities for Research in Astronomy (AURA), and the Department of Energy under Contract No. DE-AC02-76SF00515 with the SLAC National Accelerator Laboratory managed by Stanford University. The project leading to these results also received funding from “la Caixa” Foundation (ID 100010434), under the agreement LCF/BQ/EU21/11890114. This research made use of Stanford University computational resources.
    
    We thank Lynne Jones, Zeljko Ivezic and HyeYun Park for helpful discussions. G.M. thanks the LSSTC Data Science Fellowship Program, which is funded by LSSTC, NSF Cybertraining Grant \#1829740, the Brinson Foundation, and the Moore Foundation; his participation in the program has benefited this work.
\end{acknowledgements}

\facility{The Vera C. Rubin 8.4m Simonyi Survey Telescope Observatory}

\software{Batoid \citep{batoid2019},
          Galsim \citep{galsim2016}, 
          IPython \citep{Ipython2007}
          Matplotlib \citep{matplotlib2007}, 
          NumPy \citep{numpy2020}, 
          PyTorch \citep{Pytorch2019}}

\bibliography{bib}{}
\bibliographystyle{aasjournal}

\end{document}